\documentclass[prl,twocolumn,superscriptaddress]{revtex4}
\usepackage{amsmath,amssymb,mathrsfs,braket,hyperref,paralist,bm,graphicx,xcolor,soul}

\begin{document}
\title{Spin-dependent Optical Superlattice}
\author{Bing Yang}
\affiliation{Hefei National Laboratory for Physical Sciences at Microscale and Department of Modern Physics, University of Science and Technology of China, Hefei, Anhui 230026, China}
\affiliation{Physikalisches Institut, Ruprecht-Karls-Universit\"{a}t Heidelberg, Im Neuenheimer Feld 226, 69120 Heidelberg, Germany}
\author{Han-Ning Dai}
\affiliation{Hefei National Laboratory for Physical Sciences at Microscale and Department of Modern Physics, University of Science and Technology of China, Hefei, Anhui 230026, China}
\affiliation{Physikalisches Institut, Ruprecht-Karls-Universit\"{a}t Heidelberg, Im Neuenheimer Feld 226, 69120 Heidelberg, Germany}
\author{Hui Sun}
\affiliation{Hefei National Laboratory for Physical Sciences at Microscale and Department of Modern Physics, University of Science and Technology of China, Hefei, Anhui 230026, China}
\affiliation{Physikalisches Institut, Ruprecht-Karls-Universit\"{a}t Heidelberg, Im Neuenheimer Feld 226, 69120 Heidelberg, Germany}
\author{Andreas Reingruber}
\affiliation{Physikalisches Institut, Ruprecht-Karls-Universit\"{a}t Heidelberg, Im Neuenheimer Feld 226, 69120 Heidelberg, Germany}
\author{Zhen-Sheng Yuan}
\email[e-mail:]{yuanzs@ustc.edu.cn}
\affiliation{Hefei National Laboratory for Physical Sciences at Microscale and Department of Modern Physics, University of Science and Technology of China, Hefei, Anhui 230026, China}
\affiliation{Physikalisches Institut, Ruprecht-Karls-Universit\"{a}t Heidelberg, Im Neuenheimer Feld 226, 69120 Heidelberg, Germany}
\affiliation{CAS-Alibaba Quantum Computing Laboratory, Shanghai 201315, China}
\affiliation{CAS Centre for Excellence and Synergetic Innovation Centre in Quantum Information and Quantum Physics, University of Science and Technology of China, Hefei, Anhui 230026, China }
\author{Jian-Wei Pan}
\email[e-mail:]{pan@ustc.edu.cn}
\affiliation{Hefei National Laboratory for Physical Sciences at Microscale and Department of Modern Physics, University of Science and Technology of China, Hefei, Anhui 230026, China}
\affiliation{Physikalisches Institut, Ruprecht-Karls-Universit\"{a}t Heidelberg, Im Neuenheimer Feld 226, 69120 Heidelberg, Germany}
\affiliation{CAS-Alibaba Quantum Computing Laboratory, Shanghai 201315, China}
\affiliation{CAS Centre for Excellence and Synergetic Innovation Centre in Quantum Information and Quantum Physics, University of Science and Technology of China, Hefei, Anhui 230026, China }
\begin{abstract}
We propose and implement a lattice scheme for coherently manipulating atomic spins. Using the vector light shift and a superlattice structure, we demonstrate experimentally the capability on parallel spin addressing in double-wells and square plaquettes with subwavelength resolution. Quantum coherence of spin manipulations is verified through measuring atom tunneling and spin exchange dynamics. Our experiment presents a building block for engineering many-body quantum states in optical lattices for realizing quantum simulation and computation tasks.
\end{abstract}
\maketitle
Ultracold atoms in optical lattices constitute a promising system for creating multipartite entangled states \cite{Bloch:2008,Jaksch:1999}, which is an essential resource for quantum information processing \cite{Nielsen:2010,Raussendorf:2001}. As the neutral atoms prepared in the Mott insulating state consist of highly ordered quantum registers \cite{Bakr:2010,Sherson:2010}, multipartite entanglement can be generated via parallelly addressing single atoms together with two-body interactions \cite{Duan:2003,Vaucher:2008,Nielsen:2010,inaba:2014}. Following this route, sub-lattice addressing and $\sqrt{\text{SWAP}}$ operations in double-wells (DWs) were demonstrated \cite{Lee:2007,Anderlini:2007}, where the atomic spins in decoupled DW arrays were addressed by ultilizing the spin-dependent effect \cite{Lee:2007,Deutsch:1998}. However, extending these entangled pairs to a one-dimensional (1D) chain or a two-dimensional (2D) cluster remains challenging due to the lack of control over inter-well couplings\cite{Brown:2015}. In this context, a bichromatic lattice referred as ``superlattice" provides an alternative degree of freedom to connect the entangled pairs by tuning the relative lattice phase \cite{Trotzky:2008,Trotzky:2010}. Besides the $\sqrt{\text{SWAP}}$ operation in such superlattices, site-selective single-qubit addressing is further required to create multipartite cluster states for measurement-based quantum computation \cite{Raussendorf:2001,Vaucher:2008}.
\begin{figure}[!b]
{\includegraphics[width=8.7cm]{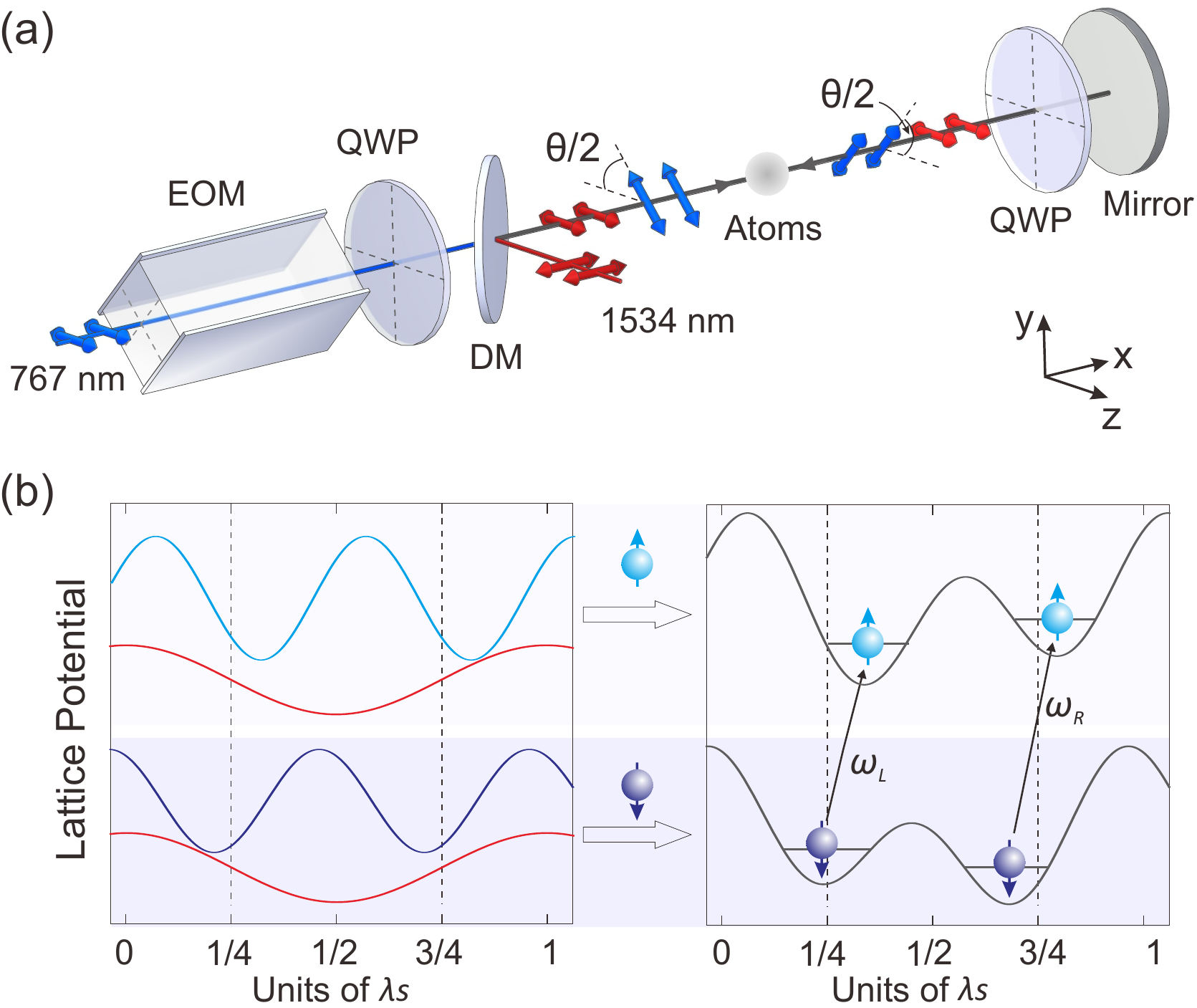}} \caption{Schematic of experimental setup and spin-dependent optical potentials. (a) The 1D superlattice is formed by overlapping two lattices on a dichromatic mirror (DM). The incident polarization of the short-lattice is controlled by an Electro-optical modulator (EOM) and a quarter wave plate (QWP), The polarization of the reflected laser is in mirror symmetry with that of the incident beam with respect to the $x-z$ plane. (b) The DWs are formed by a combination of short- and long-lattice potentials. Polarized short-lattice leads to the shift of the potential minimum in opposite directions for $\Ket{\uparrow}$ and $\Ket{\downarrow}$. The long-lattice provides a period doubling potential and breaks the symmetry of the transition frequency in the DWs. Here the optical lattices in $y$ and $z$ directions are not shown.} \label{fig1}
\end{figure}

In this Letter, we demonstrate a spin-dependent optical superlattice for coherently addressing and manipulating atomic spins. Such a lattice configuration allows one to first address even/odd rows of spins in parallel and then create entangled pairs, as well as to enable further connection of the pairs to form a multipartite entangled state with high fidelity. This configuration offers an efficient way for spin addressing in higher dimensions \cite{Dai:2016} and meanwhile becomes a powerful tool in detecting the quantum correlations of entangled states \cite{Dai:2016a}.

The optical lattice consists of two far-detuned lasers, one generates a local effective magnetic gradient, and the other one isolates the system into DWs and forms imbalanced structures for different spin states. To illustrate the spin-dependent optical potential, we consider an alkali atom placed inside a far-detuned laser field \cite{Deutsch:1998,Grimm:2000,Patrick:2013}. The monochromatic light field has a complex notation $\bm{E}(\bm{x},t) = \vec{E}(\bm{x})\exp(-\text{i}\omega t) + c.c$, where $\vec{E}$ represents the positive-frequency part with the driving frequency $\omega$. The optical potential for the atom in the ground state reads $V(\bm{x}) = - \vec{E}^\ast(\bm{x}) \cdot \bm{\alpha} \cdot \vec{E}(\bm{x})$. Here $\bm{\alpha}$ is the polarizability tensor with the irreducible scalar $\alpha_s$ and the vector components $\alpha_v$ that contribute to the potential herein \cite{Deutsch:1998}. For $^{87}$Rb with states indexed by the hyperfine $F$ and magnetic $m_F$ quantum numbers, the optical potential is a combination of scalar and vector light shift, $V = V_s +V_v$. The scalar part $V_s = - \alpha_s |\vec{E}|^2$ is state independent and proportional to the laser intensity. The vector part $V_v = \text{i} \alpha_v (\vec{E}^{\ast}\times\vec{E}) \cdot \vec{F}$ is state-dependent and can be regarded as an effective Zeeman shift with $\vec{B}_{\text{eff}} \propto \text{i} (\vec{E}^{\ast}\times\vec{E})$. This vector potential depends on the laser polarization, quantization axis and projection of the angular momentum $\vec{F}$. It vanishes for linearly polarized light or for $m_F = 0$ magnetic sublevels.

Our one-dimensional superlattice is formed by superimposing two optical standing waves differing in the period by a factor of two (see Fig. \ref{fig1}). The lattices are respectively marked as ``short-lattice" and ``long-lattice" by their wavelengths $\lambda_s$ and $\lambda_l$. Without any vector light shift, the optical dipole potential can be written as, $V(x) = V_{s} \cos^2\left( kx \right) - V_{l} \cos^2\left(kx/2 + \varphi\right)$, with $k = 2\pi/\lambda_s$ the wave number and $\varphi$ the relative phase between the lattices. The relative phase $\varphi$ is controlled by tuning the laser frequency of the long-lattice. The spin dependence arising from the laser polarization is controlled by an Electro-optical modulator (EOM). Fig. 1(a) shows the setup of a superlattice that consists of a blue- ($\lambda_s = 767$ nm) and a red-detuned ($\lambda_l = 1534$ nm) lattice, where the polarization of short-lattice is denoted as ``lin-$\theta$-lin" configuration \cite{Finkelstein:1992,Taieb:1993}. The local polarization can be decomposed into $\sigma^{\pm}$ and $\pi$ components referring to the orientation of the magnetic axis. When the magnetic axis is along $x$, the optical potential has a spin-dependent short-lattice term and a scalar long-lattice term,
\begin{equation}
\begin{split}
V_{j}(x) =& V_{s,j} \left[ A^+_{j} \cos^2\left(kx + \frac{\theta}{2}\right) +  A^-_{j} \cos^2\left(kx - \frac{\theta}{2}\right) \right]\\
&- V_{l} \cos^2\left(\frac{kx}{2} + \varphi\right).\label{eq1}
\end{split}
\end{equation}
For a certain spin state $j$, the parameters $A^+$ and $A^-$ are mainly determined by the laser detuning of the short-lattice. We can define the spin states of $^{87}$Rb as $\Ket{\downarrow} \equiv \Ket{F=1, m_F=-1}$ and $\Ket{\uparrow} \equiv \Ket{F=2, m_F=-2}$. The parameters for $\Ket{\downarrow}$ state are $A^+_{\downarrow} = 0.55$ and $A^-_{\downarrow} = 0.45$, while for spin $\Ket{\uparrow}$ are $A^+_{\uparrow} = 0.40$, $A^-_{\uparrow} = 0.60$.

The spin-dependent term of Eq. \ref{eq1} equals a periodical potential $V_{s,j} \left[A_{\text{eff}} \cdot \cos^2(kx+\theta_{\text{eff}}/2) + (1-A_{\text{eff}})/2\right]$, with effective depth $A_{\text{eff}} = \sqrt{\cos^2 \theta + (A^+ - A^-)^2\sin^2 \theta}$ and phase shift $\theta_{\text{eff}} = \tan^{-1}[(A^+ - A^-)\tan\theta]$. For $\theta = \pi/2$, the effective depth acquires a minimum $A_{\text{eff}} =  |A^+ - A^-|$ and the trap bottom shifts for $\lambda_s/4$. Until now, the coupling frequencies between $\Ket{\downarrow}$ and $\Ket{\uparrow}$ for each well in the spin-dependent short-lattice are the same (see Fig. \ref{fig1}(b)-left). Interestingly, the left-right symmetry breaks as the long-lattice adds a local potential to the DW unit and creates two different coupling frequencies $\omega_L$ and $\omega_R$ as in Fig. \ref{fig1}(b)-right. The DWs have imbalanced structures and tilt along opposite directions for spin $\Ket{\downarrow}$ and $\Ket{\uparrow}$. The superlattice with spin dependence therefore provides another degree of freedom for manipulating the atoms in the left or right wells of the DWs. Meanwhile, it creates a strong effective magnetic gradient field on the order of $\sim 250$ G/cm, which can be switched on/off with a fast speed (the EOM ramping time is 500 $\mu$s) while does not induce any unwanted eddy current. The lattice residual potential at small angles $\left(1-\cos \theta\right)$ is proportional to the second-order of the bias $ \sim \theta^2$, inducing fairly small increase of the ground-band heating compared with that in the unbiased potential. Since the EOM locates before the combining of the bichromatic lasers, it does not cause an intensity imbalance of incident and reflected beams \cite{Mandel:2003a,Lee:2007}. Such a structure neither affects the long-lattice, therefore circumventing unwanted phase fluctuations.
\begin{figure}[!b]
{\includegraphics[width=8.9cm]{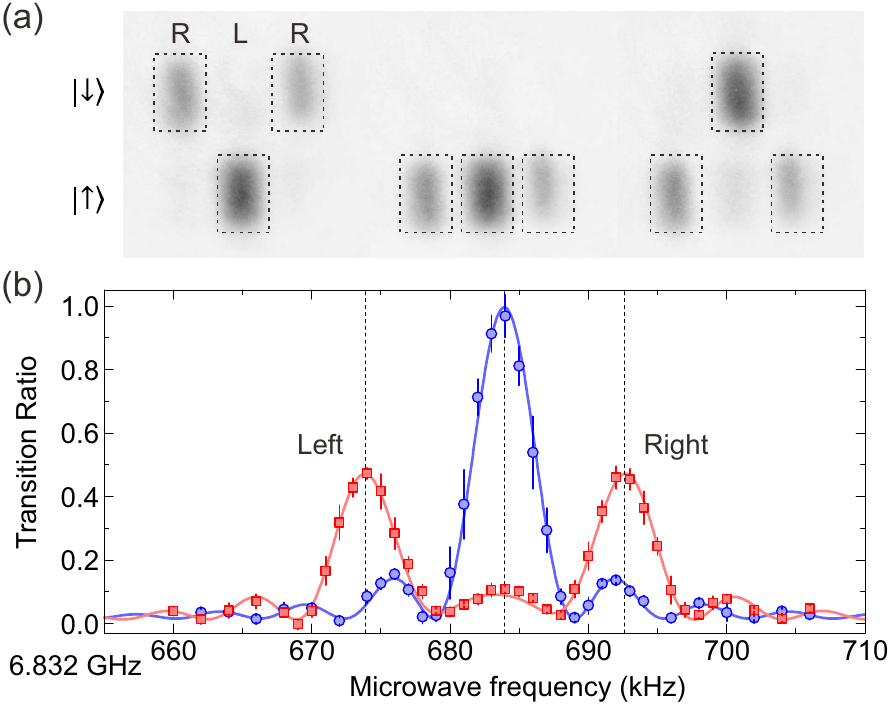}} \caption{One-dimensional spin addressing. (a) Band mapping of the spin states in DWs. Atoms in the left and right wells are mapped onto different Bloch bands. We then spatially separate the spin states by a Stern-Gerlach pulse. The left (right) picture is after spin flipping of the atoms in the left (right) sites. In the central picture, all the atoms are flipped by the MW pulse. (b) Spectroscopy of the microwave transitions. The red and blue curves are the transition ratio with and without the spin-dependent effect, respectively.   The splitting of the resonance frequency between the DW sites is 18.7(1)kHz.} \label{fig2}
\end{figure}

We implement such spin-dependent superlattice to realize the spin addressing. The experiment starts with a Bose-Einstein condensate of $^{87}$Rb with around $2 \times 10^5$ atoms in the $\Ket{F=1, m_F=-1}$ hyperfine state \cite{Dai:2016a,Dai:2016}. Then the condensate is adiabatically loaded into a single layer of a pancake-shaped trap, which is generated by interfering two laser beams with a wavelength of $\lambda_s$ and an intersection angle of $11^{\circ}$. Subsequently, we ramp up the 2D short-lattice along $x$ and $y$. As the lattice depths rise to 25 $E_r$, atoms are thereby localized into individual sites and enter an insulator state. Here $E_r = h^2/(2 m \lambda_s^2)$ denotes the recoil energy of the blue-detuned laser, with $h$ the Planck constant and $m$ the mass of atom. The $x$ and $y$ short-lattices have a frequency difference of 160 MHz to avoid the interference of these two dimensions. The filling number per lattice site and the strength of the on-site interaction are mainly controlled by adjusting the depth of the 4-$\mu$m-period ``pancake" lattice.

For the one-dimensional spin addressing in the DWs, we prepare atoms in a spin-dependent superlattice and then apply a Rabi flopping pulse on the atoms. After initialization of the insulating state, the short- and long-lattice along $x$ are ramped up to $60 E_r$ and $21 E_r$, respectively. The long-lattice divides the atoms into balanced DW units with $\varphi=0^{\circ}$. The quantization axis is set to $x$ and the phase of EOM is tuned to $\theta = 45^{\circ}$, forming an energy shift between the left and right wells as in Fig. \ref{fig1}. Subsequently, we apply a 167 $\mu$s microwave $\pi$-pulse to couple the spin $\Ket{\downarrow}$ and $\Ket{\uparrow}$. After the state addressing, the spin dependence is turned off by returning the EOM to $\theta =0^{\circ}$. Two alternative methods, site-resolved band mapping \cite{Anderlini:2007,Trotzky:2008} and $in\ situ$ imaging, are adapted to detect the spin populations (see Appendix). Fig. \ref{fig2}(a) shows the band mapping patterns of different spin states and site occupations, the efficiency for generating spin features $\Ket{\uparrow,\downarrow}$, $\Ket{\downarrow, \uparrow}$ are 92(7)$\%$, 91(5) $\%$. However, we notice that spin-exchange dynamics during the band mapping could reduce the detection fidelity \cite{Anderlini:2007,Trotzky:2008}. Since only $\Ket{\uparrow}$ react with the imaging cycling transitions, we also use \emph{in situ} absorption imaging to detect the quantum states. Fig. \ref{fig2}(b) shows the spectroscopy of site-selective addressing, where the transition peaks of the left and right DWs are separated for 18.7(1) kHz.
\begin{figure}[!htb]
{\includegraphics[width=8.3cm]{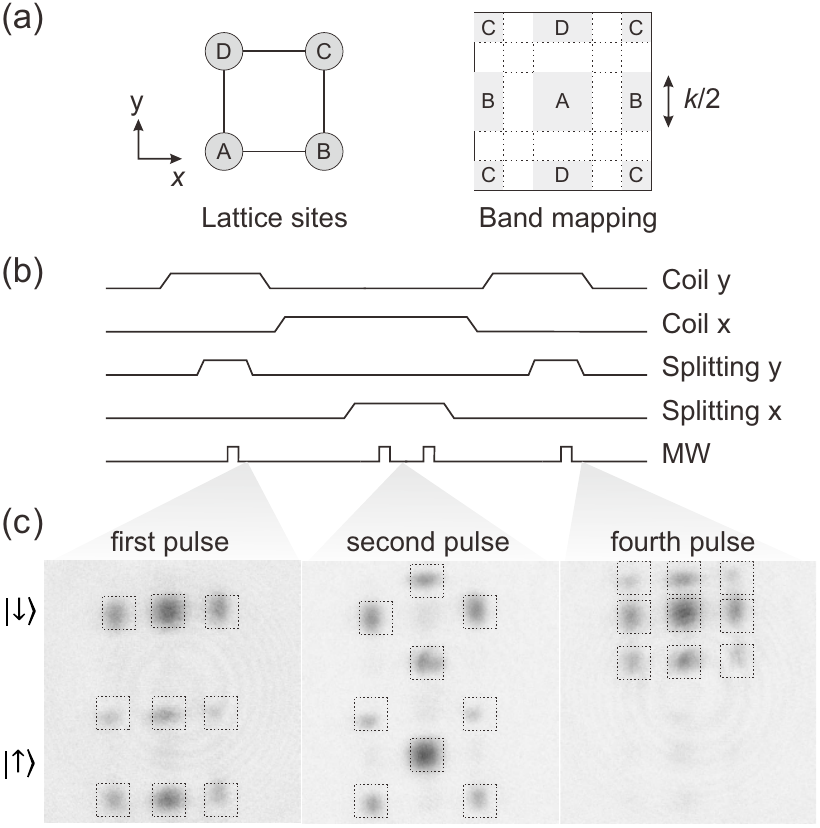}} \caption{Two-dimensional spin addressing. (a) Lattice sites are marked as A-D in each square plaquette. The band mapping pattern shows the distribution of each site in momentum space. (b) The experimental sequence of the 2D spin flipping. Before each MW pulse, the quantization axis is set and energy splitting is established. The first pulse flips the sites C and D, the second pulse flips the sites A and D, the third and fourth pulse recover the initial state. (c) Band mapping pattern after 2D spin addressing. $\Ket{\uparrow}$ and $\Ket{\downarrow}$ states are spatially separated by a Stern-Gerlach pulse. The pattern after the second MW pulse shows both the diagonal sites A and C transit to $\Ket{\uparrow}$.} \label{fig3}
\end{figure}

This spin addressing technique is extended to a two-dimensional system by implementing the superlattices on both directions. The long-lattice light along $y$ is generated by another laser source with $\sim$11GHz detuning from the frequency of $x$ long-lattice laser. Therefore, the $x$ and $y$ lattices have no crosstalk and their frequencies can be controlled individually. Atoms are initially prepared in the insulating state in the short-lattices, then the long-lattices are ramped up to form arrays of isolated square plaquettes. We perform the 2D spin addressing along each dimension in sequence, showing the capability to create a N$\acute{\text{e}}$el antiferromagnetic state inside the plaquettes. For this, the quantization axis is first set along $y$, the atoms in two sites (marked as C and D in Fig. \ref{fig3}(a)) of a plaquette are flipped by a MW $\pi$-pulse in the spin-dependent potential. Then we switch the magnetic axis to $x$ and tune the energy splitting to the desired value. A second MW $\pi$ pulse flips the spin states of sites A and D. After these two operations, atoms in the diagonal sites A and C are transferred to $\Ket{\uparrow}$, achieving a $\Ket{\uparrow,\downarrow,\uparrow,\downarrow}$ spin configuration with the site notations from A to D. For further calibrations, we then apply a third and a fourth pulse to recover the initial state by addressing $x$ and $y$ in the same fashion. Fig. \ref{fig3} shows the addressing sequences and the corresponding signals, where a 2D band mapping and Stern-Gerlach gradient analogous to the 1D case are applied to resolve the plaquette spins. The same 2D spin addressing can also be realized with a single MW pulse by setting the same transitions frequency for site A and C \cite{Dai:2016}. From the \emph{in\ situ} imaging, we calibrate the addressing efficiency of spin state $\Ket{\uparrow,\downarrow,\uparrow,\downarrow}$ to be $92(2)\%$.

\begin{figure}[!t]
{\includegraphics[width=8.6cm]{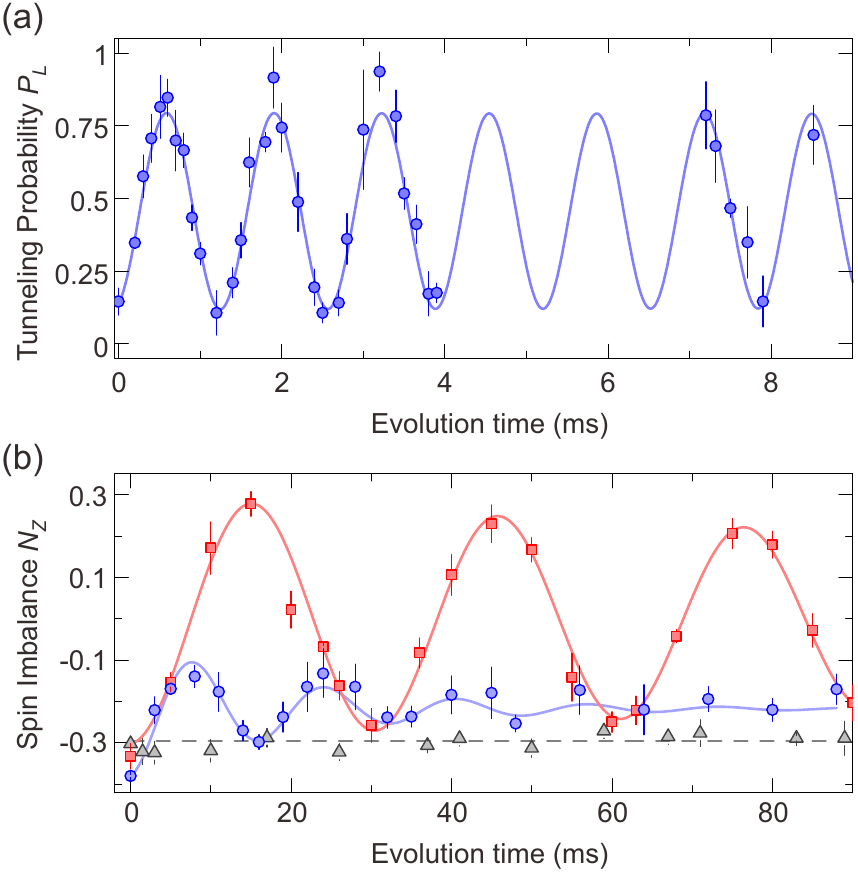}} \caption{Dynamics of (a) single atom tunneling and (b) superexchange. (a) The quantum states in DWs are initialized to $\Ket{0,\downarrow}$ by removing the atoms in the left sites. We measure the tunneling dynamics by monitoring the occupancy probability of atoms in the left sites.   The oscillation last for 6 cycles with a period of 1.32 ms. (b) Free evolution and suppressions of superexchange dynamics. The red curve is under the Hubbard parameters of tunneling $J/h =103(2)$ Hz, on-site interaction $U/h=1191(2)$Hz. While the blue and the gray curves correspond to the dynamics with extra energy shifts of $\delta/h = $ 60 Hz and $\delta/h =$ 300 Hz, respectively. The error bars represent $\pm 1\sigma$ standard deviation.} \label{fig4}
\end{figure}

To demonstrate the coherent control of spin dynamics, we study the single atom tunneling and spin-exchange process in DW systems. Atom tunneling $J$ represents the nearest-neighbor hopping term in the Bose-Hubbard model \cite{Jaksch:1999}. The spin-exchange dynamics is driven by a second-order interaction described with the Heisenberg spin model $\hat{H} = -2J_{\text{ex}} \hat{S}_L \cdot \hat{S}_R$ \cite{Duan:2003,Trotzky:2008,Dai:2016a}. By isolating the atoms into DW units, we reduce the Hilbert space close to two-level systems and thereby observe the evolutions.

The starting point of the experiment is a Mott insulator state with near unity filling. To observe the single atom tunneling, the atoms in the left sites are flipped and removed from the lattices, resulting in single fillings in the right sites and zero fillings in the left sites. We then lower down the barrier of DWs and let the system evolve. After a time $t$ the spin states are frozen by ramping up the DW barriers and we address the left sites again. The signal of the following absorption imaging has only contributions from atoms which have tunnelled to the left sites, $P_L(t) =\langle \hat{n}_{\downarrow, L}\rangle$ (here $\hat{n}_{\downarrow, L}$ represents the number operator of the left site). For short- and long-lattice depths of $11.9(1)E_r$ and $10.1(1)E_r$ during the evolution, the theoretical tunneling strength $J/h = 377(8)$ Hz matches the experimental results, as can be seen in Fig. \ref{fig4}(a). The atoms are well isolated in DWs and can oscillate for 6 cycles without any discernible decay, indicating the excellent coherence of the system. Since the frequency is sensitive to the difference of the left and right well energy levels, the atom tunneling constitutes a sensitive tool to calibrate the superlattice phase $\varphi$.

In the limit of $J \ll U$ ($U$ is the on-site interaction of the Bose-Hubbard model), the spin-exchange interaction can be well described by a two-level system with interaction strength $J_{\text{ex}} = 2 J^2/U$ \cite{Duan:2003,Trotzky:2008}. To maximize the spin-exchange amplitude, the spin-dependent splitting should be minimized to keep the state $\Ket{\uparrow,\downarrow}$ and $\Ket{\downarrow, \uparrow}$ degenerate. Thus besides setting the EOM to $\theta = 0^{\circ}$, we also set the quantization axis to $z$ during spin-exchange evolutions. The lattices parameters for spin-exchange are $V_s=17.0(1) E_r$, $V_l=10.0(1) E_r$, resulting in an superexchange interaction strength of $J_{\text{ex}}/h = 17.8(8)$ Hz. We monitor the dynamics using the N$\acute{\text{e}}$el order parameter $N_z = (n_{\uparrow L} + n_{\downarrow R} - n_{\uparrow R} - n_{\downarrow L})/2$, where $n_{\uparrow,\downarrow; L,R } = \langle \hat{n}_{\uparrow,\downarrow; L,R }\rangle$ denotes the corresponding quantum mechanical expectation values. Fig. \ref{fig4}(b) shows such spin-exchange oscillation with a high contrast. On the other hand, the spin-dependent effect can be used to suppress the spin-exchange process by breaking the degeneracy of spin states $\Ket{\uparrow,\downarrow}$ and $\Ket{\downarrow, \uparrow}$. When the quantisation axis is set to the lattice direction, controlling the EOM can induce an energy shift $\delta$ between these two spin states. As shown in Fig. \ref{fig4}(b), the oscillations between states $\Ket{\uparrow,\downarrow}$ and $\Ket{\downarrow, \uparrow}$ are dramatically suppressed as $\delta > J_{\text{ex}}$. Such phenomena can be explained by a detuned Rabi oscillation in a two-level system, where the frequency becomes larger $\sqrt{4J_{\text{ex}}^2 + \delta^2}$ and the amplitude becomes smaller 4$J_{\text{ex}}^2/\left(4J_{\text{ex}}^2 + \delta^2\right)$. The suppression of lower order dynamics can be used to explore some high-order spin interactions, like the four-body ring-exchange interactions \cite{Dai:2016}. Another important feature is the flexibility to tune the phase of entangled Bell states via controlling this energy bias $\delta$ \cite{Dai:2016a}.

In summary, we developed a spin-dependent optical superlattice for tailoring the atomic states and spin interactions. Such a lattice provides a platform for engineering quantum states in two dimensions and detecting spin correlations with $in\ situ$ imaging, e.g., generating Bell states and observing four-body ring-exchange interactions \cite{Dai:2016a,Dai:2016}. With the capability of spin addressing and manipulation, one can explore various quantum many-body models, such as spin interactions \cite{Trotzky:2008,Paredes:2008}, artificial gauge fields \cite{Fabrice:2010,Aidelsburger:2015,Li:2016,Li:2017} and out-of-equilibrium dynamics \cite{Schreiber:2015} with an novel approach in quantum state initializations and detections. Moreover, our spin-dependent lattice could also be used in atom cooling \cite{Schachenmayer:2015,Kantian:2016}, which offers intriguing prospects for future researches on spin models and quantum magnetism \cite{Lewenstein:2012,Parker:2013,Greif:2013}.

This work has been supported by the MOST (2016YFA0301600), the NNSFC (91221204, 91421305) and the Chinese Academy of Sciences.

\section{Appendix: Site-resolved band mapping in superlattice}
To detect the atom and spin population on different lattice sites, we utilize a site-resolved band mapping sequence \cite{Anderlini:2007,Trotzky:2008} in the optical superlattice. After the state preparation, the barriers between the intra-DWs are ramped up to freeze the quantum states.We then map the atoms in the left and right DWs onto different Bloch bands and measure the occupation via absorption imaging after a time-of-flight. The superlattice phase $\varphi$ is first tuned adiabatically to 70$^{\circ}$, matching the energy of the right ground band with the highly excited band of the left site. Then the DWs are merged by lowering down the short-lattice barrier in 300 $\mu$s, whereafter the $x$ long-lattice and $y$ short-lattice are ramped down in 600 $\mu$s. Finally, we apply a magnetic gradient during the time-of-flight to separate the spins (Stern-Gerlach separation), mapping out the spin and site populations into different Brillouin zones. Fig. \ref{fig2}(a) shows the band mapping patterns of different spin states and site occupations. The size of the Brillouin zones along $x$ is half of the size along $y$-direction, reflecting the double lattice constant of the long-lattice.

\bibliographystyle{apsrev4-1}
\bibliography{SDSL}
\end{document}